\documentclass[12pt,preprint]{aastex}

\shorttitle{Interstellar Carbon}
\shortauthors{Sofia and Meyer}

\begin{document}

\title{Interstellar Carbon in Translucent Sightlines\footnote{Based on
observations with the NASA/ESA {\it Hubble Space Telescope}, obtained at
the Space Telescope Science Institute, which is operated by the Association
of Universities for Research in Astronomy, Inc. under NASA contract
NAS 5-26555.}}

\author{Ulysses J. Sofia\footnote{sofiauj@whitman.edu}}
\affil{Astronomy Department, Whitman College,
    Walla Walla, WA 99362}

\author{James T. Lauroesch\footnote{jtl@elvis.astro.nwu.edu}
and David M. Meyer\footnote{davemeyer@northwestern.edu}}
\affil{Department of Physics and Astronomy, Northwestern University, 
Evanston, IL 60208}

\author{Stefan I. B. Cartledge\footnote{cartleg@taurus.phys.lsu.edu}}
\affil{Department of Physics and Astronomy, Louisiana State University, 
Baton Rouge, LA 70803}

\begin{abstract}

We report interstellar C  {\sc ii} column densities or upper limits 
determined from weak absorption of the $\lambda$2325.4029 \AA\ 
intersystem transition observed in six translucent sightlines 
($A_{V} \gtrsim 1$) with STIS.  The sightlines sample a wide range of 
interstellar characteristics including total-to-selective extinction,
$R_{V} = 2.6 - 5.1$; average hydrogen density along the sightline, $<$n(H)$>$ 
$= 3 - 14$ cm$^{-3}$; and fraction of H in molecular form, 0 - $\sim 40\%$. 
Four of the sightlines, those
toward HD 37021, HD 37061, HD 147888 and HD 207198, have
interstellar gas-phase abundances that are consistent with the diffuse 
sightline ratio of $161 \pm 17$ carbon atoms in the gas per million 
hydrogen nuclei. We note that while it has a gas-phase carbon abundance
that is consistent with the other sightlines, a large fraction of the 
C {\sc ii} toward HD 37061 is in an excited state. The sightline 
toward HD 152590 has a measured interstellar gas-phase carbon abundance
that is well above the diffuse sightline average; the column density
of C in this sightline may be overestimated due to noise structure 
in the data. Toward HD 27778 we find a 3$\sigma$ abundance upper limit 
of $\leq 108$ C atoms in the gas per million H, a substantially enhanced 
depletion of C as compared to the diffuse sightline value. 
The interstellar characteristics toward HD 27778 are 
otherwise not extreme among the sample except for an unusually 
large abundance of CO molecules in the gas.
\end{abstract}


\section{Introduction}

Carbon exists in several phases in the interstellar medium: 
dust, gaseous atoms/ions and gaseous molecules.
The phase distribution of carbon is a fundamental characteristic of 
interstellar clouds that governs their physics. In the gaseous form 
carbon plays 
a crucial role in interstellar molecular chemistry and interstellar 
cloud cooling through fine structure lines. Carbon that exists in the dust 
phase, as evidenced by spectral features attributed to amorphous 
carbons \citep{dw81}, graphite \citep{dra89} and polycyclic 
aromatic hydrocarbons \citep{jlm92}, may be 
the dominant heat source for the diffuse ISM \citep{bt94} and a 
large contributor to interstellar extinction (e.g., the 2175 \AA\space 
feature; Savage 1975; Draine 1989; L\'{e}ger et al. 1989). In order 
to understand the physical processes and 
extinction properties of the ISM, we must know the distribution of 
carbon among its possible phases.

The dominant carbon ion in neutral interstellar clouds is C {\sc ii}. 
Unfortunately the abundance of this ion is extremely difficult to measure 
because it produces only very strong absorption features at 1036 and 1334
\AA\space(log $f\lambda$ = 2.088 and 2.234, respectively; Morton 2003) and 
an extremely weak intersystem feature at 2325.4029 \AA\space 
(log $f\lambda$ = -3.954; Morton 2003). 
Complex stellar continua, multiple absorbing components and saturation 
effects severely limit the accuracy of column density determinations from
strong lines (errors can easily be a factor of 2; Fitzpatrick 1997), so
direct measurement 
of the weak intersystem line is the best way to find reliable 
C~{\sc ii} abundances in the neutral ISM.

Because of the weakness of the intersystem line, only 8 absorption
measurements of the interstellar C {\sc ii} 
gas-phase abundance have been reported in the literature. 
Seven of these abundance determinations 
were for diffuse sightlines (sightlines with total $A_{V} \lesssim 1.0$; 
Hobbs, York \& Ogerle 1982; 
Cardelli et al. 1993, 1996; Sofia et al. 1997); the eighth was 
through a translucent sightline toward HD 24534 ($A_{V} = 1.44$; Sofia, 
Fitzpatrick \& Meyer 1998). If the reasonable assumption is made that
all of the gas-phase C in diffuse sightlines is in the form of C {\sc ii},
then diffuse cloud absorption
measurements indicate that the fractions of C in the
dust and gaseous-atomic phases do not vary among these regions.

\citet{dwe97} made a measurement of the C/H ratio in the cool
neutral ISM at high Galactic latitudes
using an observation of the [C  {\sc ii}]  
158 \micron\ emission with FIRAS. Making reasonable assumptions about the 
physical conditions in the emitting regions, they found an approximate 
value of C  {\sc ii}/H  {\sc i} $\approx 50 - 100 \times 10^{-6}$.
Even at the upper end of the range, the implied C/H is lower than 
any reliable absorption measurement \citep{sfm98}.
Note that the Dwek et al. measurement does not take into account any 
molecular or ionized components of the gas; they argue however that their 
observation is not likely dominated by contributions from 
these regions. The combination of using emission measurements and
an instrument with a beam resolution of 7\arcdeg\ makes it almost 
impossible to determine whether a low C abundance could be attributed
to depletions and/or to low metallicity gas. The Dwek et al. result is 
the only indication that the gas-phase C abundance may vary in the 
diffuse ISM.

Theory suggests that C in translucent clouds (individual absorbing
regions with $A_{V} \gtrsim 1$; van Dishoeck \& Black 1988) 
should be depleted by a factor of 2 to 4 more 
than in the diffuse cloud sightlines (van Dishoeck \& Black 1988,1989;
Jansen et al. 1996). 
Note that a translucent sightline (a sightline with total visual extinction
of $A_{V} \gtrsim 1$) which may be composed of several diffuse clouds, is
different than a translucent cloud. (See \citet{rac02}
 for an excellent discussion of the difficulties in determining 
whether a translucent cloud exists along a translucent sightline.) The
uncertainty in the single measured translucent sightline (toward HD 24534, 
also known as X Per) C/H abundance results in the $\pm1\sigma$ values 
being consistent with either the gas-phase C/H in diffuse clouds, or with 
C being depleted by nearly twice the diffuse cloud value \citep{sfm98}. 
\citet{rac02} 
argue that the HD 24534 sightline does not likely contain a 
translucent cloud, and at most could be composed of 50\% translucent
cloud material. 

In this paper we investigate the depletions of carbon in
translucent sightlines by expanding the observed sample beyond 
the current single measurement that is ambiguous concerning depletions. 
The goal is to find
whether the enhanced depletion expected in regions denser than diffuse 
clouds exists. In \S2 we discuss the sample and the data. Interstellar
carbon abundances are presented in \S3 and the results are discussed in
\S4.

\section{Observations and Data Reduction}

The sightlines for this study were chosen to maximize the possibility 
of finding enhanced C depletions as compared to the diffuse cloud value.
In the sample of interstellar oxygen measurements made by \citet{car01}, 
the sightlines toward HD 27778, HD 37021, HD 37061, HD 147888, 
HD 152590 and HD 207198 appeared to show 
mild to moderate enhancements of oxygen 
depletions as compared to the average diffuse cloud abundance 
\citep{mey98}. The O depletion toward HD 152590 in Cartledge et al.
was determined with respect to Kr since no H measurement was available; 
a recent $FUSE$ H determination \citep{cml03}
shows that O/H toward this star is 
actually consistent with the diffuse cloud value. A summary of the sightline
characteristics for our sample is given in Table 1.

The intrinsic weakness of the 2325 \AA\ C {\sc ii}] feature 
that is the focus of our program mandated that 
the observations be high resolution and have high 
signal-to-noise ratios (S/N $\approx$ 50 - 140 per pixel). 
To satisfy these requirements, the data were 
all obtained with the echelle mode of the STIS instrument aboard the 
{\it Hubble Space Telescope} ($\sim$ 2.5 km/s resolution). 
Four spacecraft orbits were
allocated to each target. Data for the first observed
target, HD 207198, were all
obtained through the 0 \farcs 2 x 0 \farcs 09 aperature at a central 
wavelength of 2263 \AA. The S/N of the data was below what would be 
expected from photon statistics, so the observing strategy was changed for 
the other targets. The data for HD 27778, HD 37061, HD 147888 and HD 152590
were 
obtained through the four 0 \farcs 2 x 0 \farcs 2 FP-SPLIT apertures and were
all centered at 2263 \AA; the FPA aperture observations were shorter 
than the others because of target acquisition in the first orbit. 
The possibility of a nearby contaminating star made the 
use of the FP-SPLIT apertures impractical for the HD 37021 observations. 
Instead, those data were all obtained through STIS's 0 \farcs 2 x 0 \farcs 09 
aperture but at 
several central wavelength settings; the first two orbits centered at
2263 \AA\ and one orbit each centered at 2313 and 2363 \AA.

In order to obtain the most reliable measurements of the weak carbon line, 
the data were treated independently at the three institutions represented 
by the authors: LSU, Northwestern and Whitman. For the LSU analysis, 
the data used were those calibrated and delivered by the STScI pipeline
just after each observation. The Northwestern and Whitman analyses were
based on spectra delivered by the STScI archive at the end of October 2003
with on-the-fly calibrations. We recalibrated some spectra at Northwestern 
using the latest STSDAS (version 3.1) varying the extraction slit width
to test whether this produced any significant change to the data; 
it did not. Each institution independently performed a weighted coaddition
of the individual observations for each target 
with either IRAF or IDL software.
The final spectra were nearly indistinguishable from each other with no
difference being larger than a minor fraction of the formal uncertainty 
associated with any data point. In order to assess whether any individual
observation was introducing noise structure in the region of
C {\sc ii}] absorption, we coadded the data for each
target in all combinations of 2 and 3 exposures. No anomalous noise from a
single observation was detected above the level of uncertainties in the data. 
Each of the independent data 
reduction procedures found that for the lower-quality observations
(S/N $\lesssim 100$) structure that could not be associated
with spectral features persisted in the data. This noise structure is almost 
exclusively in a downward direction and mimics small absorption features 
(of up to about 1.5 m\AA) in the spectra. The highest-quality data
(those for HD 27778 and HD 37061) appear to have much less of this
structure with the noise looking nearly statistical. 

The analyses of the data were performed independently at each of
the three author-institutions. 
Analyses included continuum fitting the spectra, 
measuring the equivalent widths of absorption features and determining the 
error associated with the equivalent width measurements. The 
formal error budgets included both statistical and continuum placement 
uncertainties; uncertainties in the oscillator strengths were not included. 
The stellar spectra were all quite well behaved in the 
region of the C {\sc ii}] absorption, so the continua across the features 
were easily fit with a low-order polynomial. 
Since the noise structure mentioned above
could contaminate the small absorption features being 
measured, the oxygen profiles (from Cycle 8 STIS observations) 
were used to define the limits over which the C 
profiles were integrated to find equivalent widths.
Figure 1 shows the normalized C (solid line) and O (dotted line) absorption 
features 
for each sightline superimposed on different vertical scales. The effects 
of the structured 
noise can be seen in at least one of the C {\sc ii}] lines; the 
HD 37021 spectrum's C {\sc ii}] profile seems to show 
absorption at velocities longward of the O absorption. 
The uncertainty introduced by the noise structure is always in
the direction of increasing the apparent column density of the 
feature. Figure 1 shows that the noise structure is sparse and does not
dominate the spectra.

\section{Carbon Column Densities}

The interstellar gas-phase carbon along the target sightlines was expected
to be primarily in the form of ground-state C {\sc ii}.  No higher ion 
states of carbon were expected to reside in the neutral gas since 
C {\sc ii}'s 
ionization energy is well above hydrogen's, but some carbon could exist 
as neutral C and/or in molecules. Each of the sightlines was previously 
observed with STIS at wavelengths that sampled C {\sc i} and the dominant 
carbon-containing molecule, CO. These observations showed that, as expected, 
CO and C {\sc i} are generally minor reservoirs of C. The sightline
toward HD 27778 has N(CO) = $2.5 \times 
10^{16}$ cm$^{-2}$ \citep{fed94,jos86}; this is the largest CO
column density among our six sightlines by about a factor of 10. 
Other C-bearing
molecules can be ignored because their abundances are insignificant 
compared to CO.
For example, CH, CN, C$_{2}$ and C$_{3}$ 
toward HD 27778 contribute a total C column density of only 
$1.8 \times 10^{14}$ cm$^{-2}$ \citep{oka03}. None of the sightlines
contains an appreciable amount of neutral C; each has a
N(C {\sc i}) that is  less than $5 \times 10^{15}$ cm$^{-2}$.

The final
source that can contribute significantly to the total gas-phase C abundance
is excited C {\sc ii}. HD 37061 is the only sightline for which an 
excited C {\sc ii}] absorption feature was detected. Two features
out of the 63.42 cm$^{-1}$ energy level, at 2326.1126 \AA\ and 2328.8374 \AA\,
were detected and measured to have equivalent widths of 
$1.01 \pm 0.26$ and $0.89 \pm 0.21$ m\AA, respectively. 
The equivalent widths for these lines 
are nearly equal although the oscillator strengths reported in Morton 
(2003) suggest that the $\lambda2326$ feature should be approximately 
twice the strength of $\lambda2328$. This leads to a large uncertainty 
in the calculated C {\sc ii}$^{*}$ abundance; 
the weighted average of the two features gives a column density of 
N(C {\sc ii}$^{*}) = 4.65 \pm 0.84 \times 10^{17}$. This 
column density is consistent
with the crude estimate possible for the C {\sc ii}$^{*} 
\lambda 1335$ damping wings. This suggests that 
toward HD 37061 approximately 35\% of the gas-phase C {\sc ii} is in an
excited state (see Table 2).

Table 2 gives a summary of the interstellar gas-phase carbon measurements 
and column densities for each sightline in this study.
The C {\sc ii}] equivalent widths and resulting column densities are the 
averages from the three independent analyses. The reported 1$\sigma$
uncertainties are the largest from the three independent determinations. 
The three measurements for each sightline all agreed with each other within
the reported 1$\sigma$ uncertainties. 
The weakness
of the C {\sc ii}] $\lambda2325$ transition allows abundances for this
ion to be determined directly from the weak line limit. We use the oscillator 
strength reported in Morton (2003; log f$\lambda = -3.954$).
The lower limit of 0.5 m\AA\ reported in Table 2
for the equivalent width in the HD 27778 spectrum is approximately a 
3$\sigma$ limit. To check that such a feature would be identifiable
in the spectrum, we produced a synthetic absorption feature of 0.5 m\AA\ 
and the width of the interstellar O {\sc i} absorption; the synthetic
feature was easily visible giving us confidence that our limit is 
conservative.

In order
to consolidate all of the C {\sc ii} absorption measurements using 
uniform atomic constants, we give a summary of previous interstellar 
gas-phase carbon measurements updated for the Morton (2003) oscillator 
strength and \citet{cml03} H abundances when appropriate, in Table 3. 
The equivalent width measurements for the listed
sightlines can found in the references except that for $\xi$ Per. We 
have remeasured the GHRS data for this spectrum and find an equivalent width 
of $0.74 \pm 0.19$ m\AA\ (the previous measurement was $1.41 \pm 0.41$;
Cardelli et al. 1993).

\section{Discussion}

We summarize the C/H abundances from Tables 2 and 3 in Figure 2. We 
plot the abundances as a function of average H density along the sightlines, 
$<$n(H)$>$, a crude proxy for density. Although $<$n(H)$>$ seems like
it would give little insight into cloud densities along sightlines, 
it shows the tightest correlations with depletions among the typical
density proxies (e.g., f(H$_{2}$), A$_{V}$, E(B-V); \citet{hb84,jen87,sno02,
car02}).  
Note that the C/H values plotted for four of our measurements have an 
arrow for the 
lower uncertainty. This is due to the fact that we cannot assess the 
possible contribution from structured noise in those data. As noted in \S 2,
this noise always acts in the direction of increasing the C abundance.
The fact that the noise is not prevalent in the spectra leads us to
believe that most of the measurements are not contaminated.

We redetermine the average diffuse-sightline gas-phase C/H abundance using 
previous absorption measurements, the updated f-value for the C {\sc ii}] 
$\lambda 2325$\AA\ in Morton (2003), and the updated equivalent width 
for the $\xi$ Per absorption (see \S 3). The weighted average for the
diffuse sightlines gives $161 \pm 17$ atoms of
C in the gas per million H. This gas-phase abundance is larger than
previously reported values, thereby indicating less C in the dust than
previously thought. If the solar C abundance of Allende Prieto, 
Lambert \& Asplund (2002; C/H = $245 \pm 23 \times 10^{-6}$) is adopted 
as the reference for the total interstellar medium abundance, then 
fewer than 100 C atoms 
per million H atoms could be incorporated into dust in diffuse sightlines.
This low dust-phase carbon abundance is problematic for extinction models 
(Snow \& Witt 1995 and references within), but can be alleviated somewhat 
if proto-solar abundances (Lodders 2003; $288 \pm 27$ C atoms per million H) 
or young F and G star abundances (Sofia \& Meyer 2001; $358
\pm 82$ C per million H) better represent the total (gas + dust)
interstellar abundance.

Four of the translucent sightlines observed for this study, toward HD 37021,
HD 37061, HD 147888 and HD 207198, have measured
C-to-H ratios that are consistent with the diffuse cloud C/H abundance.
This may not be surprising given 
that translucent sightlines have been found to be composed primarily
of diffuse cloud gas \citep{rac02}. The sightlines, however, were chosen 
for their enhanced depletions of oxygen, so one might expect the gas-phase
C/O
ratios to be higher than for the diffuse sightlines. This is the case for
these four sightlines that have a weighted average C/O = $0.68 \pm 0.08$ 
compared to the diffuse cloud weighted average of $0.47 \pm 0.05$. The
high C/O ratios in these translucent sightlines contrast 
the measurement in the translucent
sightline toward HD 24534 where C/O = $0.28 \pm 0.09$ \citep{sfm98,sno98}. 
One must keep in mind that the noise structure discussed in \S2 could affect
some of our measurements in the direction of increasing C/O.

The sightline toward HD 152590 has a measured C/H that is well above any 
other interstellar measurement, and above any standard abundances used to 
represent the total interstellar C abundance. The C {\sc ii}] and O 
{\sc i} absorption features for this sightline (see Figure 1)
look very similar
with the profiles showing only a slight deviation toward the long-wavelength 
side of the features. One possibility is that structured noise is 
particularly strong in the absorption feature and the C abundance is
highly overestimated. Another explanation, that would also explain the 
similarity of the 2 absorption profiles, is that the sightline
is passing through an atypical region of the interstellar medium with 
unusual abundances. This seems unlikely since the O/H ratio toward the 
star is not atypical \citep{cml03}. This star was reported
to have an enhanced depletion of O compared to diffuse sightlines 
by \citet{car01}. That determination was based on
the O/Kr ratio, however a recent measurement of the H abundance from $FUSE$
data shows that the apparent depletion enhancement was the result of a 
high Kr/H \citep{cm97} rather than a low O/H \citep{cml03}. While only a
2$\sigma$ result, the apparent higher than average 
Kr/H ratio toward the star leaves open the possibility that the
sightline may possess unusual abundances.

The HD 27778 sightline has a well-determined lower limit on its carbon 
column density that gives a carbon-to-hydrogen ratio of
 $\leq$ 108 C atoms per million H in the
gas. This abundance suggests a substantial enhancement in the depletion 
of C as compared to diffuse sightlines. With a gas-phase
C/O ratio of $<0.37$, it also implies that C is more substantially 
enhanced in depletion than O. Looking at the standard parameters used
to determine sightline characteristics in Table 1, there is no 
obvious reason why the HD 27778 sightline should have such a
low carbon abundance as compared to the others. In fact, it has
an $R_{V}$ that is closer to the Galactic mean than any other in our sample
(i.e., its extinction curve is close to the Galactic average;
the extinction curve is shown in \citet{jos86}), 
it has the lowest H column density, and its $<$n(H)$>$ is a very typical 
value. 
The sightline is relatively short (225 pc; Perryman et al. 1997), so
we would not expect any metallicity gradients to be affecting the abundance.
The only extreme interstellar characteristic shown in Table 1 is that,
of the sample, the HD 27778 sightline has the highest fraction of its H in 
the form of H$_{2}$. This fraction, however, is not unusual with the
translucent sightline toward HD 24534 \citep{sno98} and the diffuse sightlines
toward $\zeta$ Per and $\zeta$ Oph having substantially
larger fractions of their H in the form of H$_{2}$.
In addition to carbon and oxygen, interstellar abundances of 
Mg, Mn, Ni, Cu and Ge have been determined for the HD 27778 line of sight
\citep{car02}. When compared with sightlines of similar mean density, that
toward HD 27778 is only unusual in that the gas-phase Ni abundance is 
$\sim 40$\% of that along the typical interstellar path. 
The molecule CO has a highly unusual abundance toward HD 27778 with 
N(CO) = $2.5 
\times 10^{16}$ cm$^{-2}$ \citep{fed94,jos86}. This high CO column density 
suggests that $\gtrsim$ 10\% of the 
gas-phase C along the sightline is in the form of CO, and that the 
CO/H$_{2}$ ratio is $\sim$ 10 times larger than for any of the other 
sightlines in Table 2. The only other sightline from the entire sample
that approaches this CO abundance is that toward HD 24534. The measured
N(CO) $= 1.4 \times 10^{16}$ \citep{she02}, however, still puts the HD 24534 
sightline CO/H$_{2}$ a factor of 2.8 below the HD 27778 value and suggests
that only $\sim 4\%$ of the gas-phase C is in the form of CO.
The unusual CO abundances toward HD 27778 are 
particularly difficult to understand given 
that no more than $\sim$ \onethird\ of the material along 
the sightline can be 
in a translucent-like component \citep{rac02}.
Other than the interstellar gas-phase C, CO and Ni abundances, there 
appears to be no indication from 
observations or chemical modeling (e.g., \citet{fed94}) that the HD
27778 sightline is unusual.  
Although we do not put forth an explanation for 
it, the fact that two carbon-based characteristics, C/H and CO/H$_{2}$, 
are highly atypical toward HD 27778 suggests that they might be linked.
Understanding that link should provide information about the conditions 
under which carbon depletes beyond the diffuse-cloud level.

\acknowledgments
We thank the referee for his thoughtful comments. 
This work was supported by grants from the Space Telescope Science 
Institute to Whitman College and Northwestern University. This research has
made use of the SIMBAD database, operated at CDS, Strasbourg, France.

\clearpage





\clearpage

\begin{deluxetable}{lcccccr}
\tablecaption{Sightline characteristics. \label{tbl-1}}
\tablewidth{0pt}
\tablehead{ 
\multicolumn{3}{c}{} & \colhead{N(H)} & \colhead{Log $<$n(H)$>$} & 
\multicolumn{2}{c}
{}\\
\colhead{Star} & \colhead{E(B-V)} & \colhead{R$_{V}$} & 
\colhead{($10^{21}$ cm$^{-2}$)} & \colhead{(cm$^{-3}$)} & 
\colhead{log f(H$_{2}$)} & \colhead{References\tablenotemark{a}}
}
\startdata
HD 27778 & 0.36 & 2.9 & $2.3 \pm 0.4$ & 0.43 & -0.35 & 1,2,3  \\

HD 37021 & 0.54 & 4.6 & $4.8 \pm 1.1$ & 0.49 & \nodata\tablenotemark{b} 
& 1,2,4 \\

HD 37061 & 0.45 & 5.1 & $5.4 \pm 1.1$ & 0.45 & \nodata\tablenotemark{b} 
& 1,2,5 \\

HD 147888 & 0.52 & 4.1 & $5.9 \pm 0.9$ & 1.07 & -0.90 &  1,2,5\\

HD 152590 & 0.39 & \nodata & $2.3 \pm 0.3$ & -0.28 & -0.78 & 1,2 \\

HD 207198 & 0.59 & 2.6 & $4.8 \pm 1.1$ & 0.27 & -0.54 & 1,2,6 \\

\enddata

\tablenotetext{a}{References: (1) \citet{car01}, (2) \citet{cml03}, (3)
\citet{whi01}, (4) \citet{drr03}, (5) \citet{ccm89},
(6) \citet{aie88}.}
\tablenotetext{b}{\citet{car01} argue that very little H$_{2}$ 
exists along this sightline.} 

\end{deluxetable}

\newpage

\begin{deluxetable}{lccccc}
\tablewidth{0pt}
\tablecaption{Carbon abundances.}
\tablehead{
\colhead{} & \multicolumn{2}{c}{C {\sc ii}} & \colhead{Other C} & \colhead{} &
\colhead{} \\
\cline{2-3}
\colhead{Star}  & \colhead{W$_{\lambda}$(m\AA)}& 
\colhead{N ($10^{17}$ cm$^{-2}$)} & \colhead{N 
($10^{17}$ cm$^{-2}$)} & \colhead{C/O\tablenotemark{a}} & 
\colhead{$10^{6}$ C/H} 
}
\startdata

HD 27778  & $<0.5$\tablenotemark{b}          & $<2.19$         & 
0.29\tablenotemark{c}           & $<0.37$ & $<108$ \\

HD 37021  & $1.50 \pm 0.51$ & $6.56 \pm 2.23$ & 
0.00\tablenotemark{d}                        & $0.53 \pm 0.18$ & $137 \pm 56$ \\

HD 37061  & $2.01 \pm 0.30$ & $8.79 \pm 1.31$ & 
$4.65 \pm 0.84$\tablenotemark{e} & $0.80 \pm 0.14$ & $249 \pm 64$ \\

HD 147888 & $2.28 \pm 0.40$ & $9.97 \pm 1.75$ & 
0.02\tablenotemark{f}           & $0.66 \pm 0.12$ & $169 \pm 38$ \\

HD 152590 & $3.70 \pm 0.76$ & $16.2 \pm 3.3 $ & 
0.01\tablenotemark{f}           & $1.41 \pm 0.34$ & $558 \pm 130$ \\

HD 207198 & $2.21 \pm 0.61$ & $9.66 \pm 2.67$ & 
0.06\tablenotemark{c}           & $0.71 \pm 0.21$ & $203 \pm 72$ \\

\enddata
\tablenotetext{a}{O abundances are from \citet{car01}.}
\tablenotetext{b}{3$\sigma$ limit.}
\tablenotetext{c}{N(CO) from \citet{fed94} and N(C {\sc i}) from 
this paper.}
\tablenotetext{d}{Total for N(CO) and N(C {\sc i}) is less than $10^{15}$ 
cm$^{-2}$ (this paper)} 
\tablenotetext{e}{N(C {\sc ii}$^{*}$) from this paper.}
\tablenotetext{f}{N(C {\sc i}) from this paper.}

\end{deluxetable}

\newpage

\begin{deluxetable}{lcccclr}
\tabletypesize{\scriptsize}
\tablewidth{0pt}
\tablecaption{Previously measured carbon abundances.}
\tablehead{
\colhead{} & \colhead{N(C)} & \multicolumn{2}{c}{} & 
\colhead{Log $<$n(H)$>$} & \multicolumn{2}{c}{}\\
\colhead{Star}  & \colhead{($10^{17}$ cm$^{-2}$)} 
& \colhead{C/O} & \colhead{$10^{6}$ C/H} & \colhead{(cm$^{-3}$)} & 
\colhead{Sightline type} & \colhead{References}\tablenotemark{a}}
\startdata

$\tau$ CMa      & $0.92 \pm 0.31$ & $0.42 \pm 0.14$ & $180 \pm 63$  & -0.96
& Diffuse & 1,2,3 \\
$\delta$ Sco    & $2.62 \pm 1.31$ & $0.48 \pm 0.25$ & $202 \pm 102$ & 0.41 
& Diffuse 
& 2,3,4 \\
$\kappa$ Ori    & $<0.79        $ & $ <0.72      $ & $ <233     $   & -0.66
& Diffuse 
& 2,3,5 \\
$\lambda$ Ori   & $1.01 \pm 0.39$ & $0.46 \pm 0.19$ & $155 \pm 65$  & -0.38
& Diffuse 
& 2,3,5 \\
$\beta^{1}$ Sco & $2.36 \pm 0.48$ & $0.56 \pm 0.15$ & $182 \pm 41$  & 0.41 
& Diffuse 
& 2,5,6 \\
$\zeta$ Oph     & $2.27 \pm 0.52$ & $0.53 \pm 0.13$ & $162 \pm 39$  & 0.53 
& Diffuse 
& 2,3,7 \\
$\zeta$ Per     & $2.23 \pm 0.39$ & $0.43 \pm 0.09$ & $139 \pm 30$  & 0.11 
& Diffuse 
& 2,3,5 \\
$\xi$ Per       & $3.23 \pm 0.83$ & $0.50 \pm 0.15$ & $170 \pm 47$  & 0.05
& Diffuse 
& 2,6,8 \\
FIRAS high latitude  &  \nodata & \nodata    & 
$\approx 50-100$\tablenotemark{b}  & \nodata & Diffuse & 9 \\
HD 24534           & $3.41 \pm 0.96$ & $0.28 \pm 0.09$ & $155 \pm 45$  & 
0.03 & Translucent & 6,10,11,12 \\
HD 154368       & $ <8.6       $ & $ <0.72      $ & $  <206      $  & 
0.40 & Translucent & 6,11,13 \\

\enddata
\tablenotetext{a}{References: (1) \citet{sof97}, (2) \citet{car01}, (3)
\citet{cml03}, (4) \citet{hyo82}, (5) \citet{car96}, (6) \citet{per97},
(7) \citet{car93}, 
(8) This paper, (9) \citet{dwe97} (10) \citet {sfm98}, (11) \citet{sno98}, 
(12) \citet{rac02}, (13) \citet{sno96}.}
\tablenotetext{b}{Values listed as an approximation in Dwek et al. (1997)}

\end{deluxetable}



\newpage

\plotone{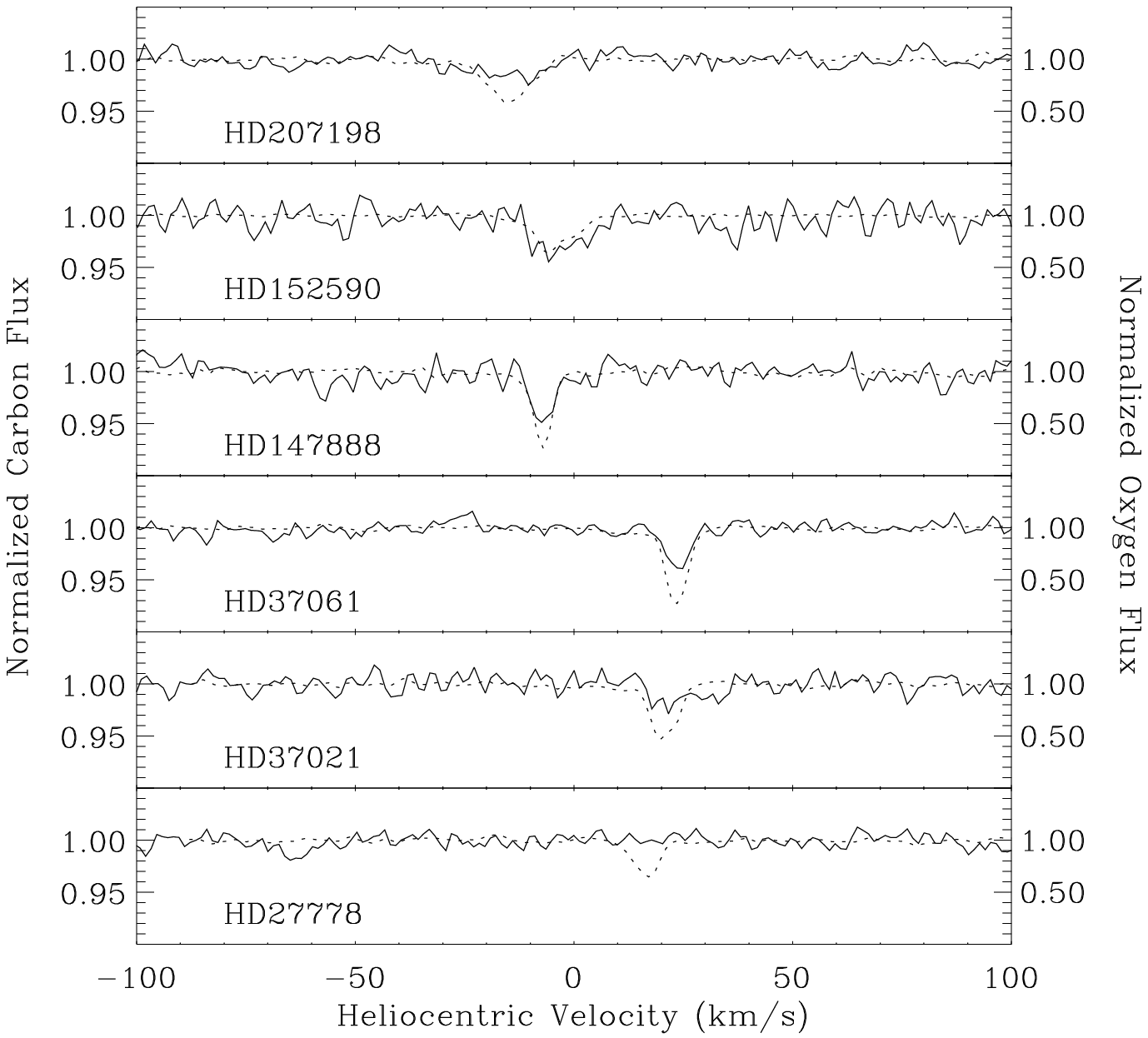}

\figcaption{Normalized STIS echelle spectra of the C {\sc ii}] 
$\lambda$2325 (solid line) and O {\sc i} $\lambda$1356 (dotted line)
absorption features. The spectra displayed are the results of the LSU data 
reduction procedure. Note that the normalized flux scale is on the 
left for carbon and the right for oxygen.}

\plotone{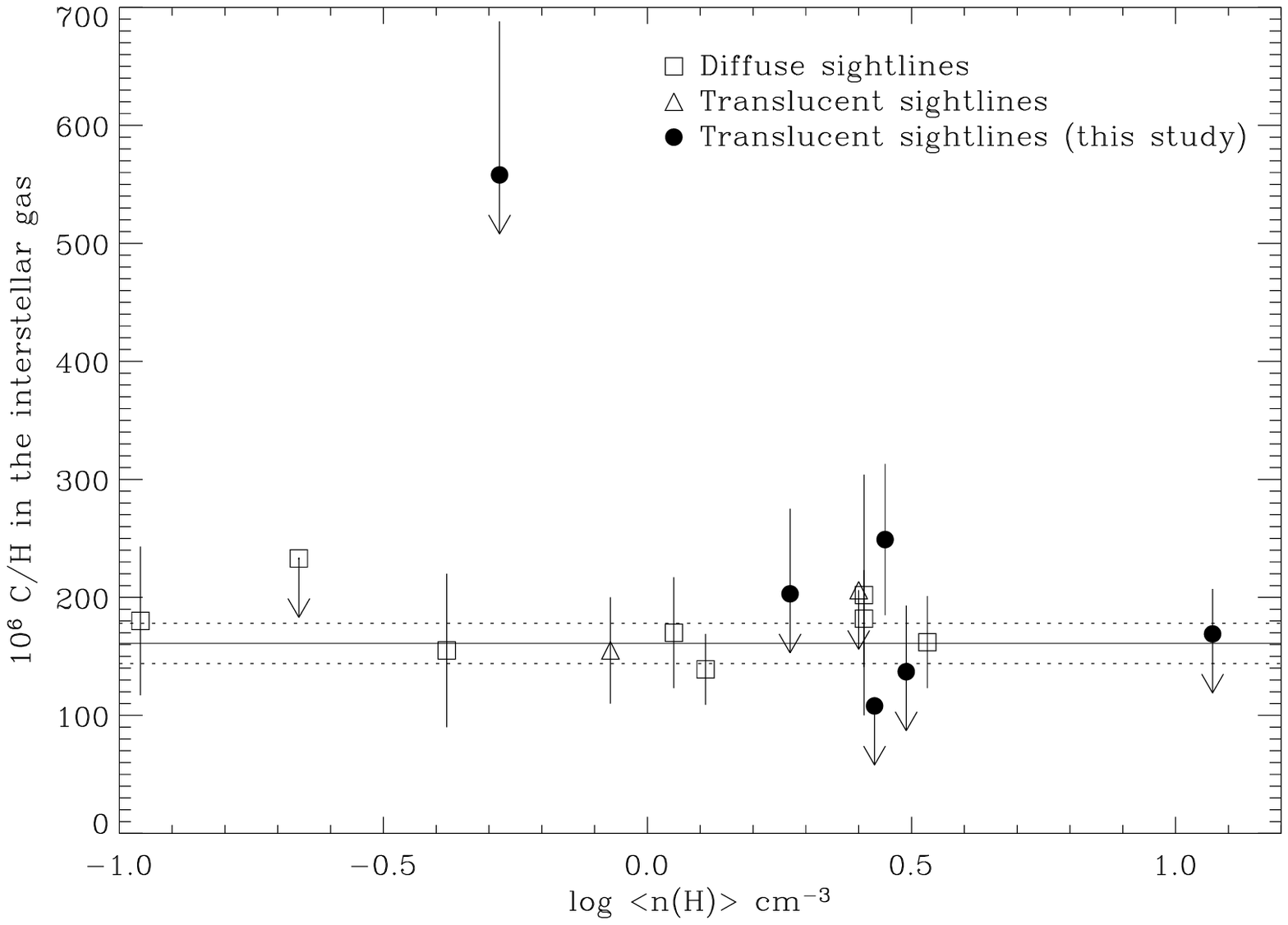}

\figcaption{Interstellar gas-phase carbon abundance measurements or limits
as a function of sightline average H density.
The solid dots are the translucent sample from this paper; the error
bars represent 1$\sigma$ uncertainties, and the arrows indicate
that the lower limit to the abundance is uncertain because of
possible noise structure in the data. The other points represent diffuse
and translucent sightlines from the literature updated for the Morton (2003)
C {\sc ii}] f-value and \citet{cml03} when available. The \citet{dwe97}
diffuse cloud value is not plotted because the value for $<$n(H)$>$ 
is unknown.
The solid line is the diffuse
sightline average based on absorption measurements with the 1$\sigma$
uncertainties shown as dotted lines.}


\end{document}